\documentclass[letterpaper,pra,aps,twocolumn,showpacs,superscriptaddress]{revtex4}
\usepackage{bm,color}
\usepackage[T1]{fontenc}
\usepackage[utf8]{inputenc}
\usepackage{amsmath} %better view for mathematical operations

\usepackage{tikzexternal}

\newcommand{\bea}[1]{\begin{align}#1\end{align}}

\newcommand{\com}{\,\text{,}}
\newcommand{\dt}{\,\text{.}}

\renewcommand{\L}{\mathcal L}
\newcommand{\del}{\partial}
\newcommand{\D}{\mathrm{d}}
\newcommand{\dd}[2]{\frac{\mathrm{d} #1}{\mathrm{d} #2}}
\newcommand{\ddel}[2]{\frac{\del #1}{\del #2}}

\newcommand{\bbm}{\begin{pmatrix}}
\newcommand{\ebm}{\end{pmatrix}}

\newcommand{\vb}[1]{\left( #1 \right)}
\newcommand{\vsb}[1]{\left[ #1 \right]}

\newcommand{\mbf}[1]{\mathbf{#1}}

\newlength{\figurewidtha}
\newlength{\figurewidthb}
\newlength{\braclen}
\newcommand{\figref}[1]{(see Fig.~\ref{#1})}

\newcommand{\V}[2]{\ensuremath{#1 \, \mathrm{#2}}}

\newcommand{\res}{\ensuremath{_\text{res}}}

\tikzsetfigurename{figure}
\newcommand{\pgfx}[2]{\pgfmathprintnumber[fixed,fixed zerofill, precision = #1]{#2}}

\begin{document}

\title{Optimization of collisional Feshbach cooling of an ultracold nondegenerate gas}
\author{Marlon Nuske}
\affiliation{Zentrum f\"ur Optische Quantentechnologien and Institut f\"ur Laserphysik, Universit\"at Hamburg, 22761 Hamburg, Germany}
\author{Eite Tiesinga}
\affiliation{Joint Quantum Institute and Center for Quantum Information and Computer Science, National Institute of Standards and Technology and University of Maryland, Gaithersburg, Maryland 20899, USA}
\author{L. Mathey}
\affiliation{Zentrum f\"ur Optische Quantentechnologien and Institut f\"ur Laserphysik, Universit\"at Hamburg, 22761 Hamburg, Germany}

\begin{abstract} 
We optimize a collision-induced cooling process for ultracold atoms in the nondegenerate regime. 
It makes use of a Feshbach resonance, instead of rf radiation in evaporative cooling, to selectively expel hot atoms from a trap.
Using functional minimization we analytically show that for the optimal cooling process the resonance energy must be tuned such that it linearly follows the temperature.
Here, optimal cooling is defined as maximizing the phase-space density after a fixed cooling duration.
The analytical results are confirmed by numerical Monte-Carlo simulations.
In order to simulate more realistic experimental conditions, we show that background losses do not change our conclusions, while additional non-resonant two-body losses make a lower initial resonance energy with non-linear dependence on temperature preferable.
\end{abstract}

\date{\today}


\maketitle

\section{Introduction}
The development of advanced cooling techniques has been of major importance from the beginning of ultracold atom experiments. In order to observe new quantum collective phenomena it is often necessary to reach lower temperatures and higher densities than are available today. Laser cooling and slowing \cite{hansch_cooling_1975,chu_three-dimensional_1985,dalibard_dressed-atom_1985,lett_observation_1988,dalibard_laser_1989,letokhov_1995,metcalf} is the most widely used technology to get from room temperature down to a few micro kelvin. It uses light pressure and the Doppler effect to decelerate the atoms while simultaneously lowering the temperature. Temperatures that can be reached with these techniques are limited by the atom recoil from a single photon and are too high for many experiments. Hence, it is usually followed by evaporative cooling \cite{hess_1986,ketterle_1996}, which selectively removes atoms that carry the highest kinetic energy. In a magnetic trap this is achieved by applying rf radiation, and in a dipole trap by lowering the laser intensity. The remaining atoms will reach a lower temperature after collisional rethermalization. Quickly lowering the laser intensity (or quickly changing the rf frequency) results in a fast cooling process but with low remaining densities. The opposite case, while promising high densities, is often limited by other constraints in an experiment. A compromise is then used.

Advanced methods of laser cooling reaching temperatures below the recoil limit of a single photon have been reported. These include velocity selective coherent population trapping (VSCPT) \cite{aspect_laser_1988} and sub-recoil Raman cooling \cite{kasevich_laser_1992,hamann_resolved-sideband_1998}, but also cavity induced cooling of single  \cite{cirac_laser_1995,horak_cavity-induced_1997,vuletic_three-dimensional_2001,maunz_cavity_2004,zippilli_cooling_2005,leibrandt_cavity_2009} and multiple \cite{mishina_cavity_2014} atoms. More recently experiments have shown demagnetization cooling \cite{hensler_depolarisation_2005,fattori_demagnetization_2006,medley_spin_2011,volchkov_efficient_2014}
and proposals for cooling atoms by extracting entropy from the system have been made \cite{ho_universal_2009,ho_squeezing_2009,bernier_cooling_2009,heidrich-meisner_quantum_2009}.
Finally, it has also been demonstrated that it is possible to cool diatomic \cite{shuman_laser_2010} and polyatomic \cite{zeppenfeld_sisyphus_2012} molecules.  
Despite these efforts to find a more efficient cooling mechanism, evaporative cooling has remained the most widely used technology in experiment.

%% Figure  %%
\begin{figure}
\includegraphics[width=8.3cm]{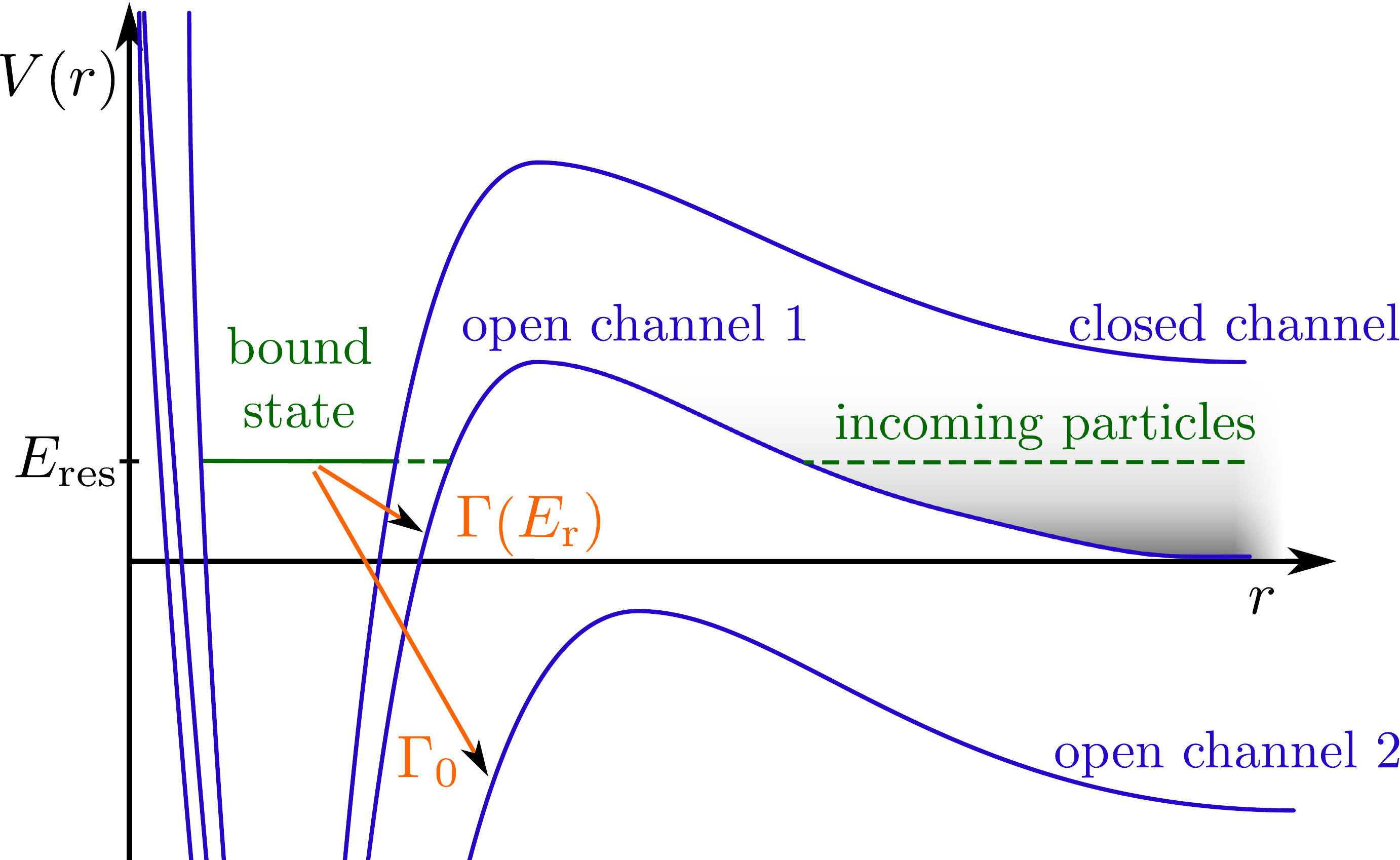}
\caption{(color online) Schematic depiction of the Feshbach cooling process for a p-wave resonance. 
The figure shows the potentials of three scattering channels as a function of atomic separation.
Scattering starts in open channel 1. The shaded band indicates the thermal distribution of the atoms. 
A semi-bound state with energy $E_{\rm res}$ exists in the closed channel and only atoms colliding with relative energy $E_{\rm r}$ close to $E_{\rm res}$ scatter significantly.
The colliding atoms return to channel 1 with rate $\Gamma(E_{\rm r})$ or go to open channel 2 with energy-independent rate $\Gamma_0$. The first process leads to thermalization of the sample while the second leads to selective loss of hot atoms.}
\label{fig:feshbach}
\end{figure}

In Ref.~\cite{mathey} we proposed to use a collisional Feshbach resonance \cite{tiesinga_1993,feshbach_1}, whose energy can be controlled with an external magnetic field, to remove hot atoms as an alternate to regular evaporative cooling. Feshbach resonances appear in a scattering process when there is a bound state with energy $E_{\rm res}$ in a closed scattering channel near the threshold energy of an open scattering channel as shown in Fig.~\ref{fig:feshbach}. 
In other words two scattering atoms with relative kinetic energy $E_{\rm r}$ temporarily form a bound state that then decays back into the original scattering channel.
In cold atom experiments Feshbach resonances are widely used to tune the scattering length of colliding atoms.
For this purpose it is crucial to not induce collisional losses and thus have only one open channel.

%%% Figure %%
\begin{figure}[b!t]
\includegraphics[width=4.3cm]{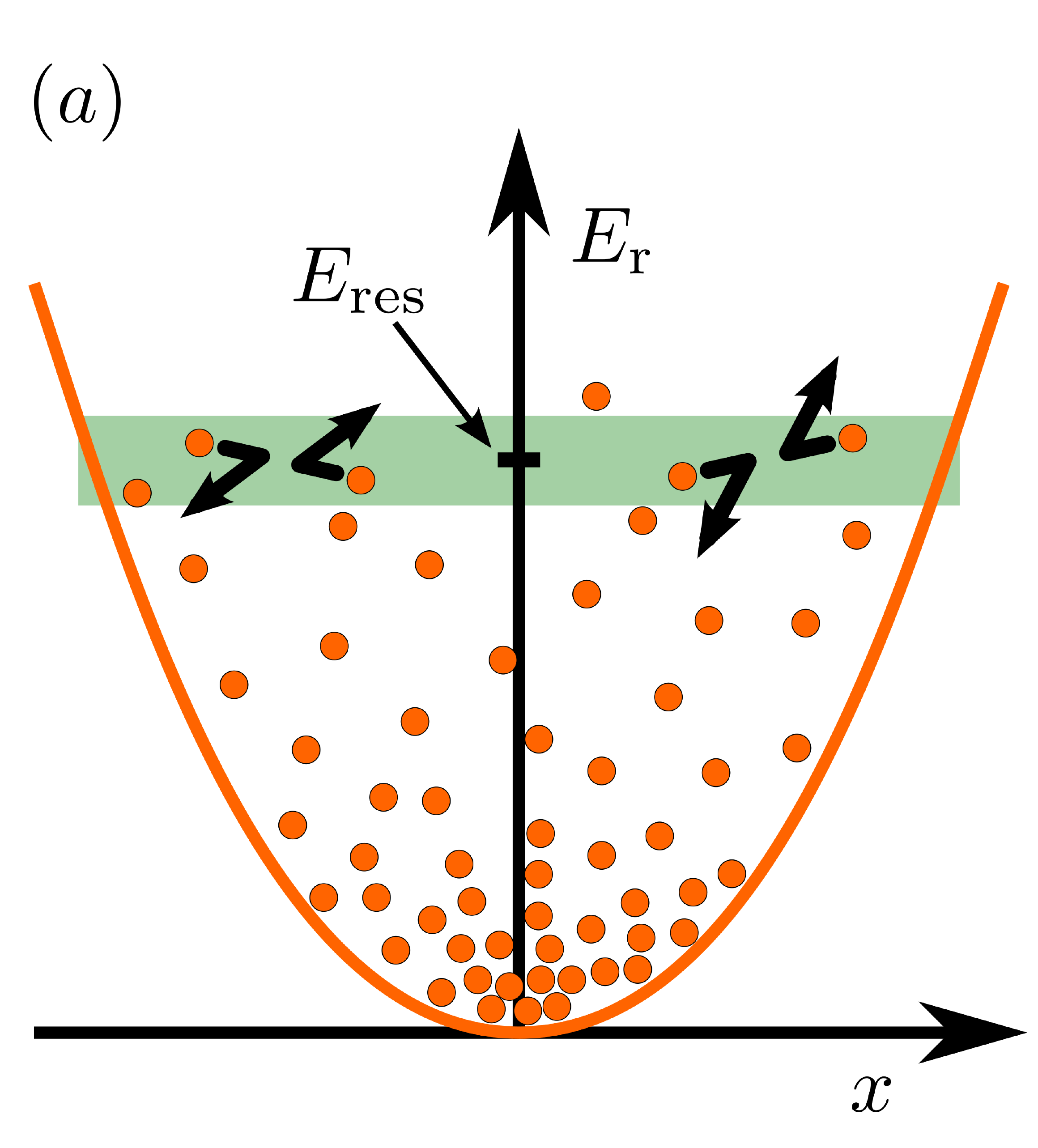}%
\includegraphics[width=4.3cm]{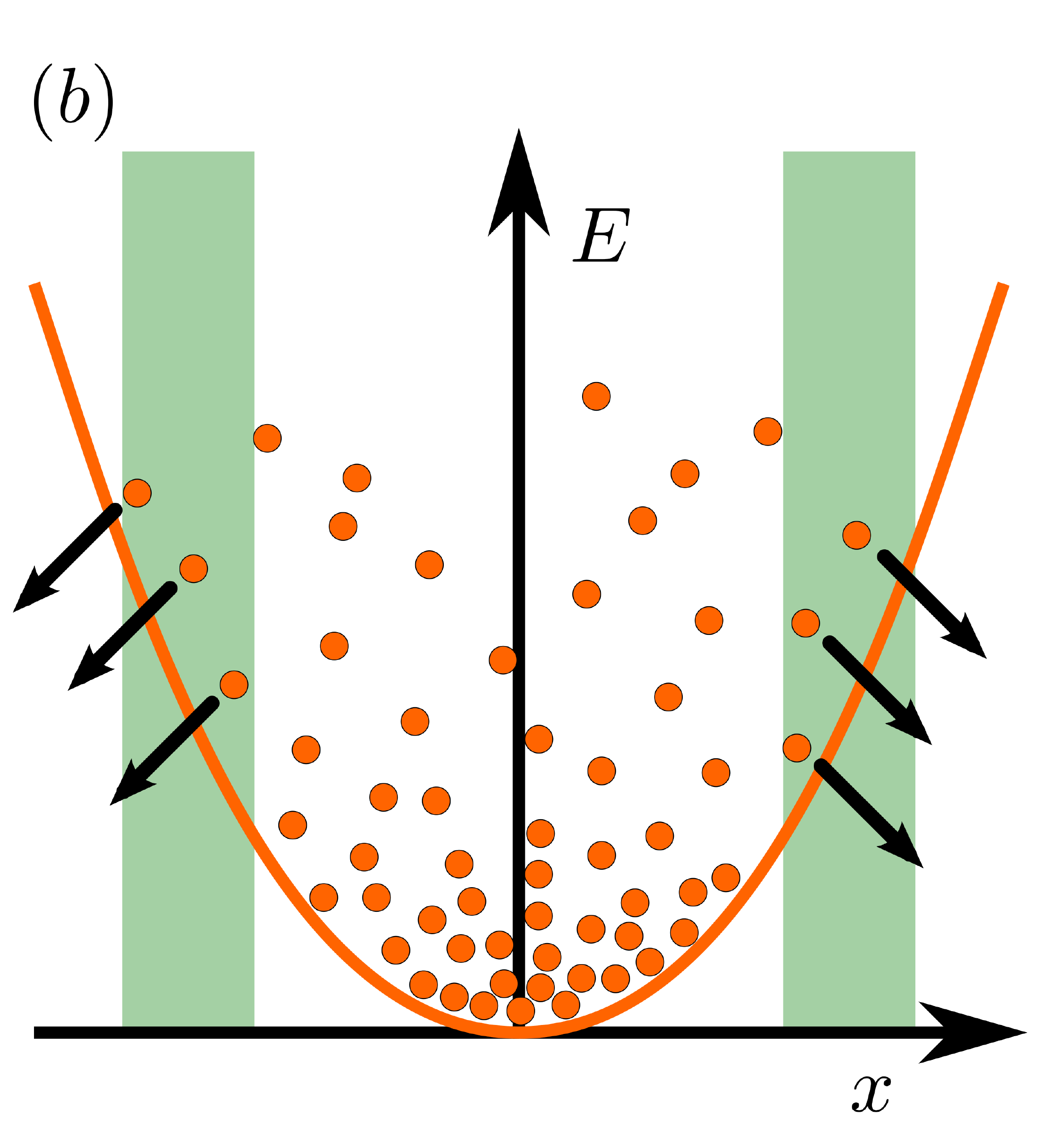}%
\caption{(color online) Comparison of the cooling mechanism using Feshbach resonances (a) and evaporative cooling in a magnetic trap (b). We sketch an ensemble of atoms in a harmonic trap (orange line) with the position along the {\it x} axis and the relative energy $E_{\rm r}$ and total energy $E$ of the atoms  along the {\it y} axis in panel (a) and (b), respectively. Atom loss, indicated by black arrows, occurs in the green shaded regions.}
\label{fig:feshbach_simp}
\end{figure}

For the Feshbach cooling process of Ref.~\cite{mathey} a ``lossy'' Feshbach resonance with a bound state that can decay into a second open scattering channel with rate $\Gamma_0$ is used \figref{fig:feshbach}. Particles scattering into this second channel are either untrapped or gain sufficient kinetic energy to be lost from the trap. Elastic scattering of particles into the initial channel with rate $\Gamma(E_{\rm r})$ leads to thermalization of the sample. The conceptual difference between Feshbach and evaporative cooling is described in Fig.~\ref{fig:feshbach_simp}. 
The first cooling process is selective in the relative kinetic energy removing atoms near the resonance location $E_{\rm res}$, while the second process only removes atoms near the edge of the cloud and, as the atoms oscillate in the trap, all hot atoms will be removed eventually.
Tuning the laser intensity or the rf frequency in evaporative cooling corresponds to changing the resonance energy in Feshbach cooling. 

The Feshbach cooling process has not yet been optimized for reaching the lowest temperature $T$ and highest density $n$ within a reasonable cooling time. 
In this paper we determine the most efficient cooling procedure for a nondegenerate homogeneous gas. Our calculations are equally valid for both bosons and fermions. We present analytical results for any ``lossy'' narrow Feshbach resonance as well as numerical results using the $^{40}$K $p$-wave resonance \cite{gaebler_2007}, which was used in Ref.~\cite{mathey} to study cooling of a spin-polarized fermionic gas.

We analytically determine the optimal cooling procedure in the framework of functional minimization. We find that the highest final phase-space density $\varpi \propto n/{T^{3/2}}$ after a fixed cooling duration is reached when the resonance energy depends linearly on temperature: $E_\text{res}(t)=(9/2)k_BT(t)$, where $k_B$ is the Boltzmann constant. This result is independent of the initial density and temperature.
We also find that when maximizing ${\cal B}= {n^a}/{T^b}$ with arbitrary non-negative coefficients $a$ and $b$ and fixed cooling duration, the phase-space density (with $a=1$ and $b=3/2$) is the only quantity, for which the optimal choice for the resonance energy, $E_\text{res}(t)=(9/2)k_BT(t)$, does not depend on the cooling duration.

We obtain the same results when adding losses due to collisions with room-temperature background molecules.
When including additional non-resonant two-body losses, however, we find that the optimal resonance energy is non-linear in temperature. 

The remainder of the paper is organized as follows. In section \ref{cool_without} we give the derivation of our analytic optimization procedure based on the phase-space density. Monte-Carlo simulations, described in section \ref{montecarlo}, confirm the analytical results. In section \ref{phase_space} we optimize with respect to other functions ${\cal B}$ instead of the phase-space density. Background loss processes and non-resonant collisional losses are included in Sec.~\ref{cool_with}. Finally, we conclude in Sec.~\ref{conclusions}.

\section{Optimization procedure}\label{cool_without}
In this section we derive the optimal time dependence for the resonance energy $E_\text{res}(t)$ in a homogeneous non-degenerate gas in order to reach the highest phase-space density $\varpi(n,T)=(2\pi\hbar)^3n/(2\pi mk_BT)^{3/2}$ after a fixed cooling duration $t_f$. Here $n$ is the density, $T$ is the temperature, $m$ is the mass of the atoms and $\hbar$ is the reduced Planck's constant.

In principle, the cooling process in a non-degenerate gas is determined by the Boltzmann equation. As in Ref.~\cite{mathey} we simplify this multi-dimensional equation by assuming that thermalization is much faster than the removal of atoms, so that the system is completely described by a Maxwell-Boltzmann distribution with a time dependent $n(t)$ and $T(t)$.
Formally, this corresponds to the case that the loss rate from the resonant state to the second open channel $\Gamma_0$ is much smaller than the rate to the initial channel $\Gamma(E_{\rm r})$.
In fact, both rates need to be much smaller than the temperature for energies $E_{\rm r}$ significantly populated by the thermal distribution, i.e. $\hbar\Gamma_0\ll \hbar \Gamma(E_{\rm r})\ll k_BT$, in order to selectively remove hot atoms. 
In this limit the momentum and spatial degrees of freedom in the Boltzmann equation can be integrated out and we obtain
\bea{
\ddel nt&=-\gamma_\text{in} n\label{eq:diff_n}\\
\ddel Tt&=-\gamma_\text{in}\vb{\frac{E_\text{res}}{3k_BT}-\frac 12}T\label{eq:diff_T}
}
and
\bea{
\gamma_\text{in}&=2\sqrt 2\, \Gamma_0 \,\varpi(n,T) \exp\vb{-\frac{E_\text{res}}{k_BT}}\dt
}

To optimize the cooling process, we follow a procedure similar to  deriving the Euler-Lagrange equations by minimizing an action under multiple constraints
\bea{
S[\mbf y]&=\int_0^{t_f} \D t \; \left[\L\vb{t,\mbf y,\mbf{\dot y}} +\sum_j \lambda_j(t) G_j\vb{t,\mbf y,\mbf{\dot y}}\right]\com \label{eq:action}
}
where $\L$ is the Lagrangian of the system and the vector of functions $\mbf y$ is time dependent\cite{kielhoefer}. The constraints are given by $G_j=0$ and the $\lambda_j(t)$ are time-dependent Lagrange multipliers, where $j$ runs from 1 to the number of constraints. 

Here, we want to maximize the phase-space density at time $t_f$, which we rewrite as
\bea{
\varpi_f=\varpi(n(t_f),T(t_f))=\int_0^{t_f}\D t \;\dd{\varpi}{t} \dt\label{eq:phase_space}
}
It is then natural to define the transposed vector $\mbf y^T = (n,T,E_\text{res})$,  
\bea{
\L(t,\mbf y, \mbf{\dot y})=- \dd \varpi t
}
and the constraints from equations \ref{eq:diff_n} and \ref{eq:diff_T}
\bea{
\bbm G_1(t,\mbf y, \mbf{\dot y}) \\ G_2(t,\mbf y, \mbf{\dot y}) \ebm=\bbm \dot n + \gamma_\text{in}(\mbf y) n\\ \displaystyle \dot T + \gamma_\text{in}(\mbf y) \vb{\frac {E_\text{res}}{3k_BT} -\frac 12} T\ebm\com \label{eq:constraints}
}
where we made explicit the functional dependence of the rate $\gamma_\text{in}$. Minimizing the action $S[\mbf y]$ by performing a functional variation with respect to $\mbf y$ gives the Euler-Lagrange equations
\bea{
\vb{\dd{}t \ddel{}{\dot y_i}-\ddel{}{y_i}} \vsb{\L\vb{t,\mbf y,\mbf{\dot y}} +\sum_j\lambda_j(t) G_j\vb{t,\mbf y,\mbf{\dot y}}}=0\label{eq:func_min}\com
}
for every $i\in \{1,2,3\}$. 

Our constraints are nonholonomic as they depend on time derivatives of the function $\mbf{ y}$. (Holonomic constraints do not depend on these derivatives.)
Consequently, the $\lambda_j$ are functions of time rather than constants. Finally, a mathematically rigorous formulation  of Eq.~\ref{eq:func_min} giving all necessary preconditions, which are indeed fulfilled in our case, can be found in ``Satz 2.7.3'' of Ref.~\cite{kielhoefer}. 

Equations \ref{eq:func_min} lead to a differential equation for the optimal $E_\text{res}(t)$. First by explicit calculation, one can see
\bea{
\dd{}{t}\vb{\ddel{\L(t,y,\dot y)}{\dot y_i}}=\ddel{\L(t,y,\dot y)}{y_i}\label{eq:cancel_l}
}
as $\L =- \D{\varpi}/ \D t$ is a total time derivative of a function with no explicit dependence on $t,\dot n, \dot T$ and $\dot E_\text{res}$. 
Hence, Eq.~\ref{eq:func_min} is independent of $\L$ and using Eq.~\ref{eq:constraints}, we obtain
\bea{
\dd{\lambda_1(t)}{t}&=\gamma_\text{in}(\mbf y)\vsb{2 \lambda_1(t)+ \frac Tn \, \vb{\frac {\eta}{3}-\frac 12}\lambda_2(t) }\label{eq:lambda1}\\
\dd{\lambda_2(t)}{t}&=\gamma_\text{in}(\mbf y)\vsb{  \frac nT  \vb{\eta-\frac 32}\lambda_1(t)+\vb{\frac{\eta^2}3-\eta+\frac 14}\lambda_2(t)}\label{eq:lambda2}\\
0&=\frac{n}{T}\lambda_1(t)+ \vb{\frac {\eta}{3}-\frac 56}\lambda_2(t)\label{eq:rel_eta}\com
}
where we define the dimensionless quantity $\eta=E_\text{res}/(k_BT)$. 
We solve Eq.~\ref{eq:rel_eta} for $\eta$, take its time derivative, insert Eqs.~\ref{eq:diff_n}, \ref{eq:diff_T}, \ref{eq:lambda1} and \ref{eq:lambda2} and find the deceptively simple differential equation
\bea{
\dd{\eta}{t}&=\frac 12 \vb{\eta-\frac 92}\label{eq:diff_x} \gamma_\text{in}
}
for the optimal choice of $\eta(t)$.
Hence, the time dependence of $\eta(t)$ decouples from the differential equations for $\lambda_1$ and $\lambda_2$.

The first-order differential equations \ref{eq:diff_n}, \ref{eq:diff_T} and \ref{eq:diff_x} define our optimal cooling path. They are uniquely specified by their initial values $n(t=0)=n_0$, $T(t=0)=T_0$ and $\eta(t=0)=\eta_0$.
Within functional minimization, however, such a path is a necessary but not sufficient condition for optimal cooling. 
The remaining task is to find for each pair $n_0$, $T_0$ the value of $\eta_0$ such that the final phase space $\varpi_f$ is maximal.

%----------------------------------------------Figure-----------------------------------------
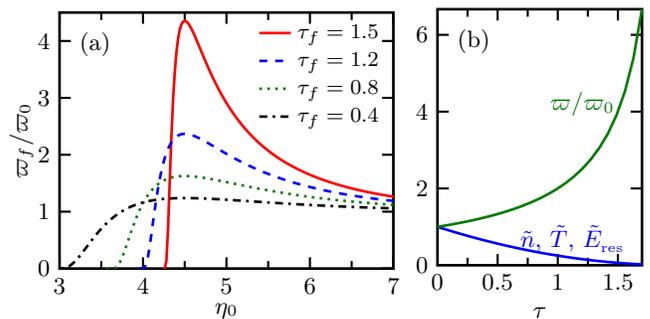
\begin{figure}[b]
\centering
 \begin{tikzpicture}[baseline]
  \begin{axis}[
	xticklabel style={/pgf/number format/.cd,fixed,precision=2},
	xtick={3,4,5,6,7,8},
	ytick={0,1,2,3,4},
	minor x tick num=1,
    	enlargelimits=false,
  	ymax=4.5,
	xmin=3,
   	xmax=7,
    	width=0.7*\figurewidtha,
    	height=0.58*\figurewidtha,
	xlabel={$\eta_0$},
    	ylabel={$\varpi_f/\varpi_0$},
	cycle list name=colorline,
	legend style={
	      	cells={anchor=west},
		anchor=north east,
		at={(1,1)},
		draw=none,
		fill=none,
		name=legendA,
		font=\footnotesize,
     	},
     	legend image post style={
     		xscale=0.7,
     	},
     	xlabel shift={-3pt},
	ylabel shift={-5pt},
	name=axisA,
]
\addplot+[no marks] table[header=true, x index=0, y expr=\thisrowno{1}]{./fesh_m40.csv};
\addlegendentry{$\tau_f=\pgfx 1{1.541}$}
\addplot+[no marks] table[header=true, x index=0, y expr=\thisrowno{1}]{./7data_ode_m40_dimless.csv};
\addlegendentry{$\tau_f=\pgfx 1{1.155}$}
\addplot+[no marks] table[header=true, x index=0, y expr=\thisrowno{1}]{./4data_ode_m40_dimless.csv};
\addlegendentry{$\tau_f=\pgfx 1{0.770}$}
\addplot+[no marks] table[header=true, x index=0, y expr=\thisrowno{1}]{./3data_ode_m40_dimless.csv};
\addlegendentry{$\tau_f=\pgfx 1{0.385}$}
\end{axis}
\node[below right=1.5mm and 1.5mm of axisA.north west]{(a)};
\end{tikzpicture}%
\begin{tikzpicture}[baseline]
\begin{axis}[
	enlargelimits=false,
	minor x tick num=1,
	ymin=0,
    	width=0.5*\figurewidtha,
    	height=0.58*\figurewidtha,
    	xlabel={$\tau$},
    	ylabel={},
    	legend style={
	      	cells={anchor=west},
     		legend pos=north west
     		},
	cycle list name=colorlinesolid,
	name=axisB,
]
\pgfplotsset{cycle list shift=1}
\addplot gnuplot[id=xoptcool1, domain=0:1.7, no markers]{(1-0.5*x)**2}
	node[pos={0.7},label={[label distance=-5pt]90:{$\tilde n$, $\tilde T$, $\tilde E_{\res}$}}] {}
;
\addplot gnuplot[id=xoptcool2, domain=0:1.7, no markers]{1/(1-0.5*x)}
	node[pos={0.6},label={[label distance=-5pt]180:{$\varpi/\varpi_0$}}] {}
;
\end{axis}
\node[below right=1.5mm and 1.5mm of axisB.north west]{(b)};
\end{tikzpicture}
   \caption{(color online) (a) Final phase-space density as a function of $\eta_0$ for several dimensionless cooling durations $\tau_f$. For all $\tau_f$ the optimal cooling procedure occurs for $\eta_0=9/2$. 
(b) Dimensionless density, temperature and phase-space density as well as scaled resonance position as a function of dimensionless time $\tau$ for the optimal cooling path $E_\text{res}=(9/2)k_BT$. For both panels the quantities on the vertical axis have been scaled with respect to their initial values.}
   \label{fig:opt_cool}
\end{figure}

Before we proceed it is beneficial to express Eqs.~\ref{eq:diff_n}, \ref{eq:diff_T} and \ref{eq:diff_x} in terms of the dimensionless density $\tilde n=n/n_0$, temperature $\tilde T=T/T_0$, initial phase-space density $\varpi_0=\varpi(n_0,T_0)$ and  time $\tau=2\sqrt 2\, e^{-9/2}\,\Gamma_0 \varpi_0\, t$. This leads to
\bea{
\ddel {\tilde n}\tau&=-  \frac{\tilde n^2}{\tilde T^{3/2}} e^{-\eta+9/2} \label{eq:diff_n2}\\
\ddel {\tilde T}\tau&=-\vb{\frac{\eta}{3}-\frac 12}\frac{\tilde n}{\tilde T^{1/2}} e^{-\eta+9/2}\label{eq:diff_T2}\\
\dd{\eta}{\tau}&=\frac 12 \vb{\eta-\frac 92}\frac{\tilde n}{\tilde T^{3/2}} e^{-\eta+9/2}\label{eq:diff_x2}\com
}
which we solve from $\tau=0$ to $\tau=\tau_f\equiv 2\sqrt 2\,e^{-9/2}\,\Gamma_0 \varpi_0 t_f$ with $\tilde n(\tau=0)=1$, $\tilde T(\tau=0)=1$ and $\eta(\tau=0)=\eta_0$. The numerical factor $e^{-9/2}$ has been included in $\tau$ for later convenience.

Figure \ref{fig:opt_cool}(a) shows the final phase-space density  as a function of $\eta_0$ for different dimensionless cooling durations $\tau_f$ after numerically solving the differential equations. For every $\tau_f$ the highest phase-space density is reached at $\eta_0=9/2$, which is the only root of Eq.~\ref{eq:diff_x2} and corresponds to a time independent solution $\eta(\tau)=9/2$. In other words the optimal cooling path does not depend on $\tau_f$.
Moreover, Eqs.~\ref{eq:diff_n2} and \ref{eq:diff_T2} can be solved analytically and we obtain
\bea{
\tilde n(\tau)=\tilde T(\tau)&=\vb{1-\tau/2}^{2}\label{eq:opt_cool_path}\text{ and }\eta(t)=9/2\dt
}
The scaled resonance position $\tilde E_{\res}(\tau)=E_\text{res}(\tau)/[(9/2)k_B T_0]$ equals $\tilde T(\tau)$ and the scaled phase-space density satisfies
\bea{
\frac{\varpi(\tau)}{\varpi_0}=\vb{1-\tau/2}^{-1}\dt
}
Figure \ref{fig:opt_cool}(b) shows the time evolution of $\tilde n$, $\tilde T$, $\tilde E_{\rm res}$ and $\varpi$ for this case. The equations also show that the phase-space density diverges at $\tau =2$. Before we reach this time, however, we must by necessity have passed the point where $\varpi(\tau)$ has become larger than one, which is beyond the validity of our classical kinetic theory with $\varpi<1$.

We have analytically verified that $\eta(t)=\eta_0=9/2$ is indeed optimal by expanding the solutions of  Eqs.~\ref{eq:diff_n2}, \ref{eq:diff_T2} and \ref{eq:diff_x2} around the optimal cooling path $\mbf x^{(0)}=(\tilde n^{(0)},\tilde T^{(0)},\eta^{(0)})$ given in Eq.~\ref{eq:opt_cool_path}. Formally, we introduce the expansion $\mbf x=(\tilde n,\tilde T,\eta)^T=\mbf x^{(0)} +\epsilon \mbf x^{(1)} + {\epsilon^2} {\bf x}^{(2)}+\cdots$ with small parameter $\epsilon$ and solve for $\mbf x^{(1)}$ and $\mbf x^{(2)}$ by Taylor expanding Eqs.~\ref{eq:diff_n2}, \ref{eq:diff_T2} and \ref{eq:diff_x2} up to second order in $\epsilon$. In fact, we find $\varpi(\tau_f)-\varpi^{(0)}_f\propto (\eta^{(0)}-9/2)^2$ with a negative proportionality constant independent of $\eta^{(0)}$.

Finally, we evaluate the cooling time and final phase-space density for the $^{40}$K Feshbach resonance characterized in \cite{gaebler_2007} with ${\hbar \Gamma_0}/{k_B}=\V{0.9}{nK}$. Starting with $n_0= \V{10^{13}}{cm}^{-3}$ and $T_0=\V{1}{\mu K}$ as in \cite{mathey} corresponding to initial phase-space density $\varpi(t=0)=0.2$, we reach $\varpi_f=1$ after a cooling duration $t_f=2.1$ s or equivalently $\tau_f=1.6$.

\section{Monte-Carlo simulations\label{montecarlo}}

%-------------------------------------------Figure----------------------------------------
\begin{figure}
\begin{center}
 \begin{tikzpicture}
  \begin{axis}[
	xticklabel style={/pgf/number format/.cd,fixed,precision=2},
	xtick={0,0.5,1.0,1.5},
	minor x tick num=1,
   	ymax=1.2,
	xmin=0,	
	xmax=1.58,
    	xlabel={$\tau$},
    	width=\figurewidthb,
    	ylabel={$\tilde E_\text{res}$},
    	legend style={
      		cells={anchor=west},
     		legend pos=south west,
    		},
	cycle list name=colorline,
	clip=false,
]
	\addplot gnuplot[id=xsteps1, domain=0:1.58, no markers]{(1-0.5*x)**2}
		node[pos={1.0},pin={[pin edge=solid,pin distance=5pt]-30:{%
		\pgfx 2{4.353382}}}
		] {}
	;
	\addlegendentry{optimal}
	\addplot+[no marks, domain=0:4,const plot] table[header=false, x expr=\thisrowno{0}*1.58/8, y expr=\thisrowno{1}/4.5e-6]{./6data_e_n_cut_0.csv}
		node[pos={1.0},pin={[name=WFn, pin edge=solid,pin distance=5pt]-30:{%
		\pgfx 2{1.46230559}}}
		] {}
	;
	\addlegendentry{initial}
 	\addplot+[no marks, domain=0:4, const plot] table[header=false, x expr=\thisrowno{0}*1.58/8, y expr=\thisrowno{1}/4.5e-6]{./6data_e_n_cut_1.csv}
 		node[pos={1.0},pin={[pin edge=solid,pin distance=5pt]-30:{%
 		\pgfx 2{1.6098646}}}
 		] {}
 	;
 	\addlegendentry{intermediate}
	\addplot+[no marks, domain=0:4,const plot] table[header=false, x expr=\thisrowno{0}*1.58/8, y expr=\thisrowno{1}/4.5e-6]{./6data_e_n_cut_4.csv}
		node[pos={1.0},pin={[pin edge=solid,pin distance=5pt]30:{%
		\pgfx 2{3.61615504}}}
		] {}
	;
	\addlegendentry{final}
	\node[above=\braclen of WFn.north] {$\overbrace{\qquad\quad}^{\textstyle \varpi_f/\varpi_0}$};
  \end{axis}
 \end{tikzpicture}
\caption{(color online) Three cooling paths as a function of $\tau$ (blue, green and black) as obtained during the Monte-Carlo procedure. The simulation assumes a piecewise-constant function for $E_{\rm res}(\tau)$. The red curve shows the optimal cooling path as obtained by functional minimization. For each cooling path we give the normalized final phase-space density to the right of the figure.}\label{fig:eight_steps}
\end{center}
\end{figure}
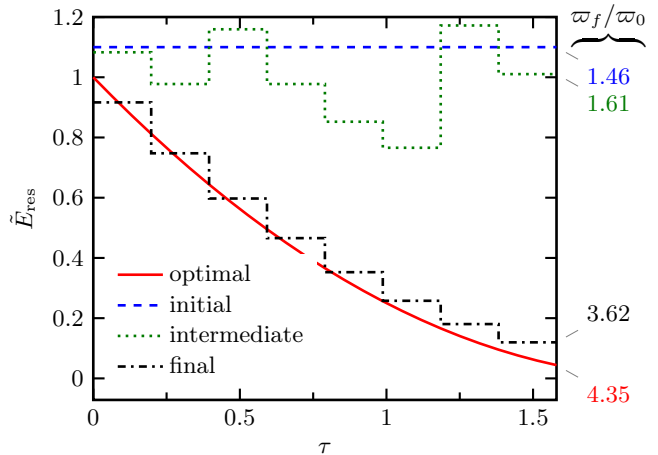

We have verified the results of section \ref{cool_without} by numerical Monte-Carlo simulations. 
Our procedure is as follows: we start with an arbitrary guess for $E_{\rm res}(t)$ and compute the final phase-space density $\varpi_f$ by solving Eqs.~\ref{eq:diff_n} and \ref{eq:diff_T} in our dimensionless quantities $\tau$, $\tilde n$, $\tilde T$ and $\tilde E_{\rm res}(\tau)=E_\text{res}(\tau)/[(9/2)k_B T_0]$. The shape of $\tilde E_{\rm res}(\tau)$ is then changed and the differential equations are solved to find a new final phase-space density $\varpi_f'$. The  shape is always accepted if $\varpi_f'>\varpi_f$ or accepted with probability $p(\varpi_f'-\varpi_f)<1$, if $\varpi_f'<\varpi_f$. This procedure is repeated for a fixed number of steps storing the current optimal cooling trajectory.

For a practical implementation we choose $\tilde E_{\rm res}(\tau)$ to be a piecewise-constant function on $L$ equal time intervals between $\tau=0$ and $\tau_f$. Here, we take $L=8$ and set  $\tau_f=1.58$ such that we can still significantly increase $\varpi$, while simultaneously stay within the classical limit $\varpi_f<1$, where our theory is valid. Furthermore, we start with an initial $\tilde E_{\rm res}$ equal to 1.1, run for $10^7$ steps and at each step allow random changes according to a Gaussian distribution with standard deviation $0.004$. Finally, we use $p(x)=\exp(500x)$ for $x<0$.
%centered around the previous value

The results of our Monte-Carlo simulation are shown in Fig.~\ref{fig:eight_steps}. It can be observed that the resonance energy $E_\text{res}(t)$ converges towards the analytically determined optimal value, which is given by $(9/2)k_BT(t)$. Additional numerical calculations assuming a polynomial function for $E_\text{res}(t)$ and optimizing its coefficients with the Monte-Carlo procedure match well with the above results.

\section{Importance of the phase-space density}\label{phase_space}
A high value of the phase-space density $\varpi$ is necessary for the observation of many quantum phenomena,
such as the Bose-Einstein phase transition at $\varpi = 2.612$. 
To examine the importance of the phase-space density in the Feshbach cooling process we set aside our goal of maximizing the final phase-space density $\varpi_f$ and instead maximize a function ${\cal B}= {n^a}/{T^b}$, with arbitrary coefficients $a$ and $b$, after a fixed dimensionless cooling duration $\tau_f$. 
As we will show the phase-space density is the only function of this type where the optimal cooling path does not depend on $\tau_f$.

First note that Eq.~\ref{eq:cancel_l} holds for the Lagrangian $\L=-\D{{\cal B}(n,T,E_\text{res})}/\D t$ since ${\cal B}$ has no explicit dependence on $t\text{, } \dot n\text{, } \dot T$ and $\dot E_\text{res}$. As $\L$ then cancels out of the calculation, Eqs.~\ref{eq:diff_n2}, \ref{eq:diff_T2} and \ref{eq:diff_x2} remain valid.

%-----------------------------------------Figure-------------------------------------------
\begin{figure}[tb]
{\footnotesize
 \begin{tikzpicture}[baseline]
  \begin{axis}[
	xticklabel style={/pgf/number format/.cd,fixed,precision=2},
	minor x tick num=1,
    	enlargelimits=false,
  	ymax=24,
	xmin=2.5,
   	xmax=5.5,
    	width=0.5\figurewidtha,
    	height=0.7*\figurewidtha,
	xlabel={$\eta_0$},
    	ylabel={$1/\tilde T(\tau_f)$},
	cycle list name=colorline,
	legend style={
	      	cells={anchor=west},
		anchor=north west,
		at={(0,0.85)},
		draw=none,
		fill=none,
		name=legendA,
		font=\footnotesize,
     	},
     	xlabel shift={-3pt},
	ylabel shift={-5pt},
	axis on top,
	name=axisA,
]
\pgfplotsset{cycle list shift=1}
\addplot+[no marks] table[header=true, x index=0, y expr=\thisrowno{2}]{./5data_ode_m40_0.csv};
\addlegendentry{$\pgfx 1{0.2}$}
\addplot+[no marks] table[header=true, x index=0, y expr=\thisrowno{2}]{./6data_ode_m40_0.csv};
\addlegendentry{$\pgfx 1{0.8}$}
\pgfplotsset{cycle list shift=-2}
\addplot+[no marks] table[header=true, x index=0, y expr=\thisrowno{2}]{./8data_ode_m40_0.csv};
\addlegendentry{$\pgfx 1{1.4}$}
\addplot[blue, mark=o, mark size=3pt] coordinates {(3.214419,1.509995)};
\addplot[mgreen, mark=triangle, mark size=3pt] coordinates {(3.9232,4.659206)};
\addplot[red, mark=diamond, mark size=3pt] coordinates {(4.267,22.67983)};
\end{axis}
%add tf above the legend
\node[name=l,above right=-2 mm and -1.8 mm of legendA.north east] {};
\node[name=r,above left=-2mm and 5.8mm of legendA.north east]{};
\draw [decorate,line width=1pt,decoration={brace,amplitude=2.5pt,mirror},xshift=0pt,yshift=0pt] 
(l) -- (r) node [black,midway,xshift=0cm,yshift=0.25cm]{{\footnotesize $\tau_f$}};
\node[below right=1.0mm and 1.5mm of axisA.north west]{(a)};
\end{tikzpicture}%
 \begin{tikzpicture}[baseline]
  \begin{axis}[
 	minor x tick num=1,
 	minor y tick num=1,
    	enlargelimits=false,
   	ymin=2.5,
   	ymax=5.9,
	xmin=0,
    	width=0.68\figurewidtha,
    	height=0.7\figurewidtha,
	xlabel={\small $\tau_f$},
    	ylabel={\small $\eta_0$},
    	legend style={
      		cells={anchor=west},
		anchor=south east,
		at={(1,0)},
		fill=none,
    	},
	cycle list name=colorline,
     	xlabel shift={-3pt},
	ylabel shift={-5pt},
	axis on top,
	name=axisB,
]
\pgfplotsset{cycle list shift=1}
\addplot+[no marks] table[header=true, x expr=\thisrowno{0}*0.7702933805, y index=7]{./2data_par_ode0.csv};
\addlegendentry{${\cal B}=\tilde n\,\tilde T^{-1}$}
\pgfplotsset{cycle list shift=-1}
\addplot+[no marks] table[header=true, x expr=\thisrowno{0}*0.7702933805, y index=1]{./2data_par_ode0.csv};
\addlegendentry{${\cal B}=\tilde n\,\tilde T^{-3/2}$}
\pgfplotsset{cycle list shift=0}
\addplot+[no marks] table[header=true, x expr=\thisrowno{0}*0.7702933805, y index=5]{./2data_par_ode0.csv};
\addlegendentry{${\cal B}=\tilde n \,\tilde T^{-2}$}
\addplot+[no marks] table[header=true, x expr=\thisrowno{0}*0.7702933805, y index=3]{./2data_par_ode0.csv};
\addlegendentry{${\cal B}=\tilde T^{-1}$}
\addplot[blue, mark=o, mark size=3pt] coordinates {(0.2,3.214419)};
\addplot[mgreen, mark=triangle, mark size=3pt] coordinates {(0.8,3.9232)};
\addplot[red, mark=diamond, mark size=3pt] coordinates {(1.4,4.267)};
\end{axis}
\node[below right=1.0mm and 1.5mm of axisB.north west]{(b)};
\end{tikzpicture}
\caption{(color online) (a) Dimensionless inverse temperature as a function of $\eta_0$ for several dimensionless cooling times $\tau_f$. For each $\tau_f$ a marker indicates the optimal $\eta_0$.
(b) Initial value $\eta_0$ maximizing the final value of the function ${\cal B}$ as a function of cooling duration $\tau_f$. Different curves correspond to different choices of the parameters $a$ and $b$, as indicated in the legend.
The markers on the curve for ${\cal B}=\tilde T^{-1}$ correspond to those in panel (a).}\label{fig:dep_func_tf2}
}
\end{figure}
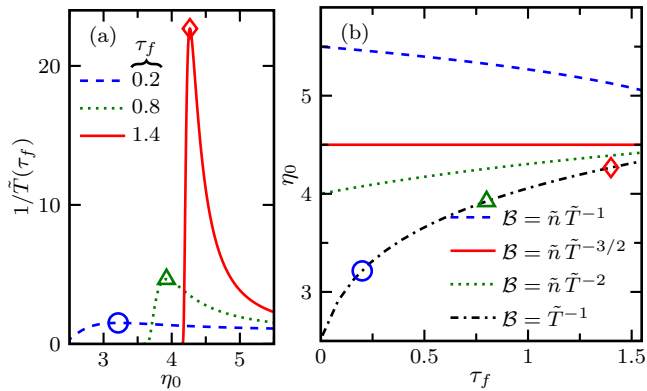

Although we obtain the same set of differential equations, which uniquely specify the optimal cooling path once $\eta_0$ has been determined, the maximal value of ${\cal B}$ does not occur at $\eta_0=9/2$ as can be seen in Fig.~\ref{fig:dep_func_tf2}(a) for the case that $a=0$ and $b=1$, corresponding to the inverse temperature $1/\tilde T$. In fact, the position of the maximum ${\cal B}$ depends on the cooling duration $\tau_f$. This is further illustrated in Fig.~\ref{fig:dep_func_tf2}(b) where  for each $\tau_f$ we plot the value of $\eta_0$ leading to the highest ${\cal B}$ for several choices of $a$ and $b$. 
We see that for all values of the parameters $a$ and $b$, except for those corresponding to $\varpi$, the optimal value of $\eta_0$ does depend on $\tau_f$.

 %----------------------------Figure--------------------------------------------
\begin{figure}[t]
{\footnotesize
\centering
 \begin{tikzpicture}[baseline]
  \begin{axis}[
	xticklabel style={/pgf/number format/.cd,fixed,precision=2},
	minor x tick num=1,
	minor y tick num=1,
    	enlargelimits=false,
 	ymax=1.9,
	xmin=3.5,
  	xmax=6,
    	width=0.6*\figurewidtha,
    	height=0.6*\figurewidtha,
	xlabel={$\eta_0$},
    	ylabel={$\tilde \varpi_f/\varpi_0$},
	cycle list name=colorline,
	legend style={
	      	cells={anchor=west},
		anchor=south east,
		at={(1,0)},
		draw=none,
		fill=none,
		name=legendA,
		font=\footnotesize,
     	},
     	legend image post style={
     		xscale=0.7,
     	},
     	xlabel shift={-3pt},
	ylabel shift={-5pt},
	name=axisA,
]
\addplot+[no marks] table[header=true, x index=0, y expr=\thisrowno{1}]{./13data_ode_m40_0.csv};
\addlegendentry{$\tau_f=\pgfx 1{0.385}$}
\addplot+[no marks] table[header=true, x index=0, y expr=\thisrowno{1}]{./14data_ode_m40_0.csv};
\addlegendentry{$\tau_f=\pgfx 1{0.770}$}
\addplot+[no marks] table[header=true, x index=0, y expr=\thisrowno{1}]{./15data_ode_m40_0.csv};
\addlegendentry{$\tau_f=\pgfx 1{1.155}$}
\addplot+[no marks] table[header=true, x index=0, y expr=\thisrowno{1}]{./16data_ode_m40_0.csv};
\addlegendentry{$\tau_f=\pgfx 1{1.541}$}
\end{axis}
\node[below right=1.5mm and 1.5mm of axisA.north west]{(a)};
\end{tikzpicture}%
 \begin{tikzpicture}[baseline]
  \begin{axis}[
	xticklabel style={/pgf/number format/.cd,fixed,precision=2},
	minor x tick num=1,
	minor y tick num=1,
    	enlargelimits=false,
   	ymin=0,
    	xmax=1.5,
    	width=0.54\figurewidtha,
    	height=0.6\figurewidtha,
	xlabel={$\tau$},
    	ylabel={},
    	legend style={
      		cells={anchor=west},
		anchor=east,
		at={(1,0.45)},
		fill=none,
		font=\footnotesize,
    	},
    	legend image post style={
	    xscale=0.6,
    	},
	cycle list name=colorline,
	xlabel shift={-3pt},
	name=axisB,
]
\pgfplotsset{cycle list shift=1}
\addplot+[no marks] table[header=true, x expr=\thisrowno{0}*exp(-4.5), y expr=\thisrowno{1}]{./16details_m40_0_ode0.csv};
\addlegendentry{$\tilde n$}
\pgfplotsset{cycle list shift =-1}
\addplot+[no marks] table[header=true, x expr=\thisrowno{0}*exp(-4.5), y expr=\thisrowno{2}]{./16details_m40_0_ode0.csv};
\addlegendentry{$\tilde T$, $\tilde E_{\rm res}$}
\pgfplotsset{cycle list shift =0}
\addplot+[no marks] table[header=true,x expr=\thisrowno{0}*exp(-4.5), y expr=\thisrowno{4}]{./16details_m40_0_ode0.csv};
\addlegendentry{$\varpi/\varpi_0$}
\end{axis}
\node[below right=1.5mm and 1.5mm of axisB.north west]{(b)};
\end{tikzpicture}
\caption{(color online) (a) Final phase-space density, including atom losses due to collisions with background atoms or molecules with rate $\gamma_\text{bg}=\V{0.2}{s^{-1}}$, as a function of the initial value of $\eta$ for several dimensionless cooling durations $\tau_f$. For all $\tau_f$ optimal cooling occurs for $\eta_0=9/2$. 
(b) Dimensionless density, temperature and phase-space density as well as scaled resonance position as a function of dimensionless time $\tau$ for the optimal cooling path $E_{\rm res}=(9/2) k_BT$ and $\gamma_\text{bg}=\V{0.2}{s^{-1}}$. 
For both panels the quantities on the vertical axis have been scaled with respect to their initial values. 
The value of $\gamma_{\rm bg}$ is a typical experimental value and the cooling procedure has noticeable loss due to collisions with background atoms and molecules, while still showing significant cooling.}\label{fig:opt_cool_with}
}
\end{figure}
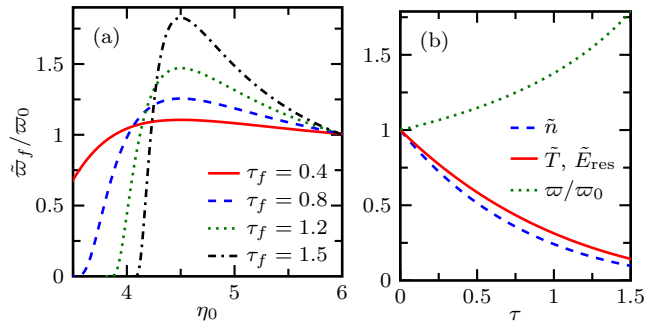
 
\section{Cooling with additional loss processes}\label{cool_with}
Processes that lead to undesired  loss of atoms from the trap are always present in experiments. These losses do not contribute to cooling and may even be heating the sample. 
One typical loss mechanism encountered in cold atomic gas experiments are collisions with atoms or molecules in the room-temperature background gas. Although operating under high vacuum conditions, background loss rates $\gamma_{\rm bg}$ may still be large enough to induce significant loss of atoms. They can be included in our theory by modifying Eq.~\ref{eq:diff_n} to
\bea{
\ddel nt&=-\gamma_\text{in} n-\gamma_\text{bg} n \dt \label{eq:dotn_bg}
}
In the functional minimization procedure optimizing the phase-space density the extra term  $-\gamma_{\rm bg}n$ drops out and once again we obtain Eq.~\ref{eq:diff_x}. By numerically solving Eqs.~\ref{eq:diff_T}, \ref{eq:diff_x} and \ref{eq:dotn_bg} in the same dimensionless units as in Sec.~\ref{cool_without} and calculating the final phase-space density for different $\eta_0$ we find that optimal cooling is again obtained for constant $\eta(t)=\eta_0=9/2$ independent of the cooling duration but also the loss rate due to background collisions. An example of such calculations is shown in Fig.~\ref{fig:opt_cool_with}(a). In fact, we find that the density and temperature are then explicitly given by
\bea{
\exp(\gamma_{\rm bg} \tau) \tilde n(\tau)=\tilde T(\tau)&=\vb{1+\frac{\exp(-\gamma_{\rm bg}\tau)-1}{2\gamma_{\rm bg}}}^{2}\dt
}
Figure \ref{fig:opt_cool_with}(b) displays the corresponding optimal cooling path
%In addition, Fig.~\ref{fig:opt_cool_with}(b) displays the optimal cooling path
for one value of $\gamma_{\rm bg}$. The additional loss process makes the dimensionless density decrease faster than the dimensionless temperature in contrast to the original case shown in Fig.~\ref{fig:opt_cool}(b). The scaled resonance position $\tilde E_{\rm res}$ is still equal to $\tilde T$.

%---------------------------------------------Figure---------------------------------------
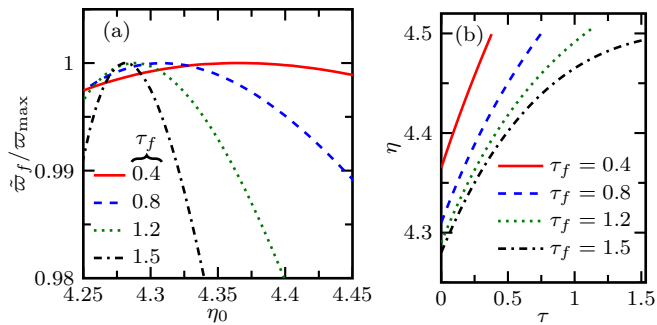
\begin{figure}[t]
{\footnotesize
\centering
 \begin{tikzpicture}[baseline]
  \begin{axis}[
	xticklabel style={/pgf/number format/.cd,fixed,precision=2},
	minor x tick num=1,
	minor y tick num=1,
    	enlargelimits=false,
 	ymin=0.98,
 	ymax=1.005,
	xmin=4.25,
  	xmax=4.45,
    	width=0.6*\figurewidtha,
    	height=0.6*\figurewidtha,
	xlabel={$\eta_0$},
    	ylabel={$\tilde \varpi_f/\varpi_{\rm max}$},
	cycle list name=colorline,
	legend style={
	      	cells={anchor=west},
		anchor=south west,
		at={(0,0)},
		draw=none,
		fill=none,
		name=legendA,
		font=\footnotesize,
     	},
     	legend image post style={
     		xscale=0.7,
     	},
     	xlabel shift={-3pt},
	ylabel shift={-5pt},
	name=axisA,
]
\addplot+[no marks] table[header=true, x index=0, y expr=\thisrowno{1}/1.130252]{./18data_ode_m40_0.csv};
\addlegendentry{$\pgfx 1{0.385}$}
\addplot+[no marks] table[header=true, x index=0, y expr=\thisrowno{1}/1.372708]{./19data_ode_m40_0.csv};
\addlegendentry{$\pgfx 1{0.770}$}
\addplot+[no marks] table[header=true, x index=0, y expr=\thisrowno{1}/1.799202]{./20data_ode_m40_0.csv};
\addlegendentry{$\pgfx 1{1.155}$}
\addplot+[no marks] table[header=true, x index=0, y expr=\thisrowno{1}/2.723558]{./21data_ode_m40_0.csv};
\addlegendentry{$\pgfx 1{1.541}$}
\end{axis}
\node[name=l,above right=-2 mm and -1.8 mm of legendA.north east] {};
\node[name=r,above left=-2mm and 5.8mm of legendA.north east]{};
\draw [decorate,line width=1pt,decoration={brace,amplitude=2.5pt,mirror},xshift=0pt,yshift=0pt] 
(l) -- (r) 
node [black,midway,xshift=0cm,yshift=0.25cm]{{\footnotesize $\tau_f$}};
\node[below right=0.5mm and 1.5mm of axisA.north west]{(a)};
\end{tikzpicture}%
%-------------------second graph----------------------------------------
\begin{tikzpicture}[baseline]
  \begin{axis}[
	xticklabel style={/pgf/number format/.cd,fixed,precision=2},
	minor x tick num=1,
	minor y tick num=1,
    	enlargelimits=false,
   	ymin=4.25,
  	ymax=4.52,
    	width=0.5\figurewidtha,
    	height=0.6\figurewidtha,
	xlabel={$\tau$},
    	ylabel={$\eta$},
    	legend style={
      		cells={anchor=west},
		anchor=south east,
		at={(1,0.05)},
		fill=none,
		font=\footnotesize,
    	},
	cycle list name=colorline,
	xlabel shift={-3pt},
 	ylabel shift={-5pt},
 	name=axisB,
]
\addplot+[no marks, restrict x to domain=0:0.385] table[header=true, x expr=\thisrowno{0}*exp(-4.5), y index=3]{./18details_m40_0_ode0.csv};
\addlegendentry{$\tau_f=\pgfx 1{0.385}$}
\addplot+[no marks, restrict x to domain=0:0.77] table[header=true, x expr=\thisrowno{0}*exp(-4.5), y index=3]{./19details_m40_0_ode0.csv};
\addlegendentry{$\tau_f=\pgfx 1{0.770}$}
\addplot+[no marks, restrict x to domain=0:1.155] table[header=true, x expr=\thisrowno{0}*exp(-4.5), y index=3]{./20details_m40_0_ode0.csv};
\addlegendentry{$\tau_f=\pgfx 1{1.155}$}
\addplot+[no marks] table[header=true, x expr=\thisrowno{0}*exp(-4.5), y index=3]{./21details_m40_0_ode0.csv};
\addlegendentry{$\tau_f=\pgfx 1{1.541}$}
\end{axis}
\node[below right=0.5mm and 0.7mm of axisB.north west]{(b)};
\end{tikzpicture}
\caption{(color online) (a) Final phase-space density, including effects due to non-resonant collisions with $n_0\gamma_2=\V{0.2}{s^{-1}}$ as shown in Eq.~\ref{eq:two_body}, as a function of $\eta_0$ for several dimensionless cooling durations $\tau_f$. In order to highlight the differences between the curves we have normalized the phase-space density with respect to the maximal value $\varpi_{\rm max}$ for each $\tau_f$. Optimal cooling occurs at smaller values of $\eta_0$ for longer $\tau_f$. %
(b) 
Time dependence of $\eta(\tau)$ for several dimensionless cooling durations $\tau_f$ and $n_0\gamma_2=\V{0.2}{s^{-1}}$ with $\eta_0$ chosen such that the phase-space density is maximized. Each curve is only plotted up to the corresponding $\tau_f$.}
\label{fig:twobody1}
}
\end{figure}

Non-resonant two-body losses are a second important loss process present in cold atom experiments. They are taken into account by replacing Eq.~\ref{eq:diff_n} with
\bea{
\ddel nt&=-\gamma_\text{in} n-\gamma_2 n^2\dt \label{eq:two_body}
}
where $\gamma_2$ is the non-resonant loss rate coefficient.
For this case the functional minimization procedure yields
\bea{
\dd{\eta}{t}&=\frac 12 \vb{\eta-\frac 92}\gamma_\text{in}+\vb{\eta-\frac 52}\gamma_2 n\com
}
instead of Eq.~\ref{eq:diff_x}. We can then numerically solve Eqs.~\ref{eq:diff_T}, \ref{eq:diff_x} and \ref{eq:two_body} in the dimensionless units from Sec.~\ref{cool_without}.
Figure \ref{fig:twobody1}(a) shows that optimal cooling occurs at different $\eta_0$ for different cooling durations. In order to compete with the additional loss process a lower initial resonance energy, corresponding to $\eta_0<9/2$, is preferable in all cases. Furthermore, Fig.~\ref{fig:twobody1}(b) shows that for every $\tau_f$ $\eta(\tau)$ rises slowly up to a final value of $\eta(\tau_f)\approx 4.5$.
This accounts for the fact that the term corresponding to non-resonant collisional losses has a quadratic dependence on density and thus, as density decreases in the cooling process, it becomes less important in comparison to $\gamma_{\rm in}n$.

In summary, the optimal choice for the resonance energy does not depend on the strength of collisions with background atoms and molecules, but has to be changed noticeably, if non-resonant losses are present.

\vspace{1cm}

\section{Conclusions}\label{conclusions}
We have analytically determined the optimal Feshbach cooling procedure using functional minimization. Keeping the resonance energy proportional to temperature $E_\text{res}(t)=(9/2)k_BT(t)$ leads to the highest phase-space density after a fixed cooling duration. We have demonstrated that the choice for the resonance energy leading to optimal cooling is independent of the initial conditions such as initial density and temperature of the atoms. Furthermore the result $E_\text{res}(t)=(9/2)k_BT(t)$  is not altered when taking into account loss of atoms due to collisions with room-temperature background atoms. When adding two body collisions, however, it is optimal to choose a resonance energy that is no longer linear in temperature. Our analytically obtained results match well with our numerical calculations.

%% acknowledgments %%
\begin{acknowledgments}
We acknowledge support from the Deutsche Forschungsgemeinschaft through the SFB 925 and the Hamburg Centre for Ultrafast Imaging, and from the Landesexzellenzinitiative Hamburg, which is supported by the Joachim Herz Stiftung. M.N. acknowledges support from the German Economy Foundation.
\end{acknowledgments}

%%%%%%%% Reference %%%%%%%%
\bibliography{bibliography}

\begin{thebibliography}{34}
\expandafter\ifx\csname natexlab\endcsname\relax\def\natexlab#1{#1}\fi
\expandafter\ifx\csname bibnamefont\endcsname\relax
  \def\bibnamefont#1{#1}\fi
\expandafter\ifx\csname bibfnamefont\endcsname\relax
  \def\bibfnamefont#1{#1}\fi
\expandafter\ifx\csname citenamefont\endcsname\relax
  \def\citenamefont#1{#1}\fi
\expandafter\ifx\csname url\endcsname\relax
  \def\url#1{\texttt{#1}}\fi
\expandafter\ifx\csname urlprefix\endcsname\relax\def\urlprefix{URL }\fi
\providecommand{\bibinfo}[2]{#2}
\providecommand{\eprint}[2][]{\url{#2}}

\bibitem[{\citenamefont{Hänsch and Schawlow}(1975)}]{hansch_cooling_1975}
\bibinfo{author}{\bibfnamefont{T.~W.} \bibnamefont{Hänsch}} \bibnamefont{and}
  \bibinfo{author}{\bibfnamefont{A.~L.} \bibnamefont{Schawlow}},
  \bibinfo{journal}{Opt.~Commun.} \textbf{\bibinfo{volume}{13}},
  \bibinfo{pages}{68} (\bibinfo{year}{1975}).

\bibitem[{\citenamefont{Chu et~al.}(1985)\citenamefont{Chu, Hollberg,
  Bjorkholm, Cable, and Ashkin}}]{chu_three-dimensional_1985}
\bibinfo{author}{\bibfnamefont{S.}~\bibnamefont{Chu}},
  \bibinfo{author}{\bibfnamefont{L.}~\bibnamefont{Hollberg}},
  \bibinfo{author}{\bibfnamefont{J.~E.} \bibnamefont{Bjorkholm}},
  \bibinfo{author}{\bibfnamefont{A.}~\bibnamefont{Cable}}, \bibnamefont{and}
  \bibinfo{author}{\bibfnamefont{A.}~\bibnamefont{Ashkin}},
  \bibinfo{journal}{Phys.~Rev.~Lett.} \textbf{\bibinfo{volume}{55}},
  \bibinfo{pages}{48} (\bibinfo{year}{1985}).

\bibitem[{\citenamefont{Dalibard and
  Cohen-Tannoudji}(1985)}]{dalibard_dressed-atom_1985}
\bibinfo{author}{\bibfnamefont{J.}~\bibnamefont{Dalibard}} \bibnamefont{and}
  \bibinfo{author}{\bibfnamefont{C.}~\bibnamefont{Cohen-Tannoudji}},
  \bibinfo{journal}{J.~Opt.~Soc.~Am.~B} \textbf{\bibinfo{volume}{2}},
  \bibinfo{pages}{1707} (\bibinfo{year}{1985}).

\bibitem[{\citenamefont{Lett et~al.}(1988)\citenamefont{Lett, Watts, Westbrook,
  Phillips, Gould, and Metcalf}}]{lett_observation_1988}
\bibinfo{author}{\bibfnamefont{P.~D.} \bibnamefont{Lett}},
  \bibinfo{author}{\bibfnamefont{R.~N.} \bibnamefont{Watts}},
  \bibinfo{author}{\bibfnamefont{C.~I.} \bibnamefont{Westbrook}},
  \bibinfo{author}{\bibfnamefont{W.~D.} \bibnamefont{Phillips}},
  \bibinfo{author}{\bibfnamefont{P.~L.} \bibnamefont{Gould}}, \bibnamefont{and}
  \bibinfo{author}{\bibfnamefont{H.~J.} \bibnamefont{Metcalf}},
  \bibinfo{journal}{Phys.~Rev.~Lett.} \textbf{\bibinfo{volume}{61}},
  \bibinfo{pages}{169} (\bibinfo{year}{1988}).

\bibitem[{\citenamefont{Dalibard and
  Cohen-Tannoudji}(1989)}]{dalibard_laser_1989}
\bibinfo{author}{\bibfnamefont{J.}~\bibnamefont{Dalibard}} \bibnamefont{and}
  \bibinfo{author}{\bibfnamefont{C.}~\bibnamefont{Cohen-Tannoudji}},
  \bibinfo{journal}{J.~Opt.~Soc.~Am.~B} \textbf{\bibinfo{volume}{6}},
  \bibinfo{pages}{2023} (\bibinfo{year}{1989}).

\bibitem[{\citenamefont{Letokhov et~al.}(1995)\citenamefont{Letokhov,
  Ol'shanii, and Ovchinnikov}}]{letokhov_1995}
\bibinfo{author}{\bibfnamefont{V.~S.} \bibnamefont{Letokhov}},
  \bibinfo{author}{\bibfnamefont{M.~A.} \bibnamefont{Ol'shanii}},
  \bibnamefont{and} \bibinfo{author}{\bibfnamefont{Y.~B.}
  \bibnamefont{Ovchinnikov}}, \bibinfo{journal}{Quantum Semiclass.~Opt.}
  \textbf{\bibinfo{volume}{7}}, \bibinfo{pages}{5} (\bibinfo{year}{1995}).

\bibitem[{\citenamefont{Metcalf and van~der Straten}(1999)}]{metcalf}
\bibinfo{author}{\bibfnamefont{H.~J.} \bibnamefont{Metcalf}} \bibnamefont{and}
  \bibinfo{author}{\bibfnamefont{P.}~\bibnamefont{van~der Straten}},
  \emph{\bibinfo{title}{Laser Cooling and Trapping}}
  (\bibinfo{publisher}{Springer}, \bibinfo{year}{1999}),
  \bibinfo{edition}{corrected edition} ed.

\bibitem[{\citenamefont{Hess}(1986)}]{hess_1986}
\bibinfo{author}{\bibfnamefont{H.}~\bibnamefont{Hess}}, \bibinfo{journal}{Phys.
  Rev. B} \textbf{\bibinfo{volume}{34}}, \bibinfo{pages}{3476}
  (\bibinfo{year}{1986}).

\bibitem[{\citenamefont{Ketterle and Druten}(1996)}]{ketterle_1996}
\bibinfo{author}{\bibfnamefont{W.}~\bibnamefont{Ketterle}} \bibnamefont{and}
  \bibinfo{author}{\bibfnamefont{N.~J.~V.} \bibnamefont{Druten}}, in
  \emph{\bibinfo{booktitle}{Adv.~At.~Mol.~Opt.~Phys.}}
  (\bibinfo{publisher}{Academic Press}, \bibinfo{year}{1996}), vol.
  \bibinfo{volume}{Volume 37}, pp. \bibinfo{pages}{181--236}.

\bibitem[{\citenamefont{Aspect et~al.}(1988)\citenamefont{Aspect, Arimondo,
  Kaiser, Vansteenkiste, and Cohen-Tannoudji}}]{aspect_laser_1988}
\bibinfo{author}{\bibfnamefont{A.}~\bibnamefont{Aspect}},
  \bibinfo{author}{\bibfnamefont{E.}~\bibnamefont{Arimondo}},
  \bibinfo{author}{\bibfnamefont{R.}~\bibnamefont{Kaiser}},
  \bibinfo{author}{\bibfnamefont{N.}~\bibnamefont{Vansteenkiste}},
  \bibnamefont{and}
  \bibinfo{author}{\bibfnamefont{C.}~\bibnamefont{Cohen-Tannoudji}},
  \bibinfo{journal}{Phys.~Rev.~Lett.} \textbf{\bibinfo{volume}{61}},
  \bibinfo{pages}{826} (\bibinfo{year}{1988}).

\bibitem[{\citenamefont{Kasevich and Chu}(1992)}]{kasevich_laser_1992}
\bibinfo{author}{\bibfnamefont{M.}~\bibnamefont{Kasevich}} \bibnamefont{and}
  \bibinfo{author}{\bibfnamefont{S.}~\bibnamefont{Chu}},
  \bibinfo{journal}{Phys.~Rev.~Lett.} \textbf{\bibinfo{volume}{69}},
  \bibinfo{pages}{1741} (\bibinfo{year}{1992}).

\bibitem[{\citenamefont{Hamann et~al.}(1998)\citenamefont{Hamann, Haycock,
  Klose, Pax, Deutsch, and Jessen}}]{hamann_resolved-sideband_1998}
\bibinfo{author}{\bibfnamefont{S.~E.} \bibnamefont{Hamann}},
  \bibinfo{author}{\bibfnamefont{D.~L.} \bibnamefont{Haycock}},
  \bibinfo{author}{\bibfnamefont{G.}~\bibnamefont{Klose}},
  \bibinfo{author}{\bibfnamefont{P.~H.} \bibnamefont{Pax}},
  \bibinfo{author}{\bibfnamefont{I.~H.} \bibnamefont{Deutsch}},
  \bibnamefont{and} \bibinfo{author}{\bibfnamefont{P.~S.}
  \bibnamefont{Jessen}}, \bibinfo{journal}{Phys.~Rev.~Lett.}
  \textbf{\bibinfo{volume}{80}}, \bibinfo{pages}{4149} (\bibinfo{year}{1998}).

\bibitem[{\citenamefont{Cirac et~al.}(1995)\citenamefont{Cirac, Lewenstein, and
  Zoller}}]{cirac_laser_1995}
\bibinfo{author}{\bibfnamefont{J.~I.} \bibnamefont{Cirac}},
  \bibinfo{author}{\bibfnamefont{M.}~\bibnamefont{Lewenstein}},
  \bibnamefont{and} \bibinfo{author}{\bibfnamefont{P.}~\bibnamefont{Zoller}},
  \bibinfo{journal}{Phys.~Rev.~A} \textbf{\bibinfo{volume}{51}},
  \bibinfo{pages}{1650} (\bibinfo{year}{1995}).

\bibitem[{\citenamefont{Horak et~al.}(1997)\citenamefont{Horak, Hechenblaikner,
  Gheri, Stecher, and Ritsch}}]{horak_cavity-induced_1997}
\bibinfo{author}{\bibfnamefont{P.}~\bibnamefont{Horak}},
  \bibinfo{author}{\bibfnamefont{G.}~\bibnamefont{Hechenblaikner}},
  \bibinfo{author}{\bibfnamefont{K.~M.} \bibnamefont{Gheri}},
  \bibinfo{author}{\bibfnamefont{H.}~\bibnamefont{Stecher}}, \bibnamefont{and}
  \bibinfo{author}{\bibfnamefont{H.}~\bibnamefont{Ritsch}},
  \bibinfo{journal}{Phys.~Rev.~Lett.} \textbf{\bibinfo{volume}{79}},
  \bibinfo{pages}{4974} (\bibinfo{year}{1997}).

\bibitem[{\citenamefont{Vuletić et~al.}(2001)\citenamefont{Vuletić, Chan, and
  Black}}]{vuletic_three-dimensional_2001}
\bibinfo{author}{\bibfnamefont{V.}~\bibnamefont{Vuletić}},
  \bibinfo{author}{\bibfnamefont{H.~W.} \bibnamefont{Chan}}, \bibnamefont{and}
  \bibinfo{author}{\bibfnamefont{A.~T.} \bibnamefont{Black}},
  \bibinfo{journal}{Phys.~Rev.~A} \textbf{\bibinfo{volume}{64}},
  \bibinfo{pages}{033405} (\bibinfo{year}{2001}).

\bibitem[{\citenamefont{Maunz et~al.}(2004)\citenamefont{Maunz, Puppe,
  Schuster, Syassen, Pinkse, and Rempe}}]{maunz_cavity_2004}
\bibinfo{author}{\bibfnamefont{P.}~\bibnamefont{Maunz}},
  \bibinfo{author}{\bibfnamefont{T.}~\bibnamefont{Puppe}},
  \bibinfo{author}{\bibfnamefont{I.}~\bibnamefont{Schuster}},
  \bibinfo{author}{\bibfnamefont{N.}~\bibnamefont{Syassen}},
  \bibinfo{author}{\bibfnamefont{P.~W.~H.} \bibnamefont{Pinkse}},
  \bibnamefont{and} \bibinfo{author}{\bibfnamefont{G.}~\bibnamefont{Rempe}},
  \bibinfo{journal}{Nature Phys.} \textbf{\bibinfo{volume}{428}},
  \bibinfo{pages}{50} (\bibinfo{year}{2004}).

\bibitem[{\citenamefont{Zippilli and Morigi}(2005)}]{zippilli_cooling_2005}
\bibinfo{author}{\bibfnamefont{S.}~\bibnamefont{Zippilli}} \bibnamefont{and}
  \bibinfo{author}{\bibfnamefont{G.}~\bibnamefont{Morigi}},
  \bibinfo{journal}{Phys.~Rev.~Lett.} \textbf{\bibinfo{volume}{95}},
  \bibinfo{pages}{143001} (\bibinfo{year}{2005}).

\bibitem[{\citenamefont{Leibrandt et~al.}(2009)\citenamefont{Leibrandt,
  Labaziewicz, Vuletić, and Chuang}}]{leibrandt_cavity_2009}
\bibinfo{author}{\bibfnamefont{D.~R.} \bibnamefont{Leibrandt}},
  \bibinfo{author}{\bibfnamefont{J.}~\bibnamefont{Labaziewicz}},
  \bibinfo{author}{\bibfnamefont{V.}~\bibnamefont{Vuletić}}, \bibnamefont{and}
  \bibinfo{author}{\bibfnamefont{I.~L.} \bibnamefont{Chuang}},
  \bibinfo{journal}{Phys.~Rev.~Lett.} \textbf{\bibinfo{volume}{103}},
  \bibinfo{pages}{103001} (\bibinfo{year}{2009}).

\bibitem[{\citenamefont{Mishina}(2014)}]{mishina_cavity_2014}
\bibinfo{author}{\bibfnamefont{O.~S.} \bibnamefont{Mishina}},
  \bibinfo{journal}{New J.~Phys.} \textbf{\bibinfo{volume}{16}},
  \bibinfo{pages}{033021} (\bibinfo{year}{2014}).

\bibitem[{\citenamefont{Hensler et~al.}(2005)\citenamefont{Hensler, Greiner,
  Stuhler, and Pfau}}]{hensler_depolarisation_2005}
\bibinfo{author}{\bibfnamefont{S.}~\bibnamefont{Hensler}},
  \bibinfo{author}{\bibfnamefont{A.}~\bibnamefont{Greiner}},
  \bibinfo{author}{\bibfnamefont{J.}~\bibnamefont{Stuhler}}, \bibnamefont{and}
  \bibinfo{author}{\bibfnamefont{T.}~\bibnamefont{Pfau}},
  \bibinfo{journal}{Europhys.~Lett.} \textbf{\bibinfo{volume}{71}},
  \bibinfo{pages}{918} (\bibinfo{year}{2005}).

\bibitem[{\citenamefont{Fattori et~al.}(2006)\citenamefont{Fattori, Koch,
  Goetz, Griesmaier, Hensler, Stuhler, and
  Pfau}}]{fattori_demagnetization_2006}
\bibinfo{author}{\bibfnamefont{M.}~\bibnamefont{Fattori}},
  \bibinfo{author}{\bibfnamefont{T.}~\bibnamefont{Koch}},
  \bibinfo{author}{\bibfnamefont{S.}~\bibnamefont{Goetz}},
  \bibinfo{author}{\bibfnamefont{A.}~\bibnamefont{Griesmaier}},
  \bibinfo{author}{\bibfnamefont{S.}~\bibnamefont{Hensler}},
  \bibinfo{author}{\bibfnamefont{J.}~\bibnamefont{Stuhler}}, \bibnamefont{and}
  \bibinfo{author}{\bibfnamefont{T.}~\bibnamefont{Pfau}},
  \bibinfo{journal}{Nature Phys.} \textbf{\bibinfo{volume}{2}},
  \bibinfo{pages}{765} (\bibinfo{year}{2006}).

\bibitem[{\citenamefont{Medley et~al.}(2011)\citenamefont{Medley, Weld, Miyake,
  Pritchard, and Ketterle}}]{medley_spin_2011}
\bibinfo{author}{\bibfnamefont{P.}~\bibnamefont{Medley}},
  \bibinfo{author}{\bibfnamefont{D.~M.} \bibnamefont{Weld}},
  \bibinfo{author}{\bibfnamefont{H.}~\bibnamefont{Miyake}},
  \bibinfo{author}{\bibfnamefont{D.~E.} \bibnamefont{Pritchard}},
  \bibnamefont{and} \bibinfo{author}{\bibfnamefont{W.}~\bibnamefont{Ketterle}},
  \bibinfo{journal}{Phys.~Rev.~Lett.} \textbf{\bibinfo{volume}{106}},
  \bibinfo{pages}{195301} (\bibinfo{year}{2011}).

\bibitem[{\citenamefont{Volchkov et~al.}(2014)\citenamefont{Volchkov, Rührig,
  Pfau, and Griesmaier}}]{volchkov_efficient_2014}
\bibinfo{author}{\bibfnamefont{V.~V.} \bibnamefont{Volchkov}},
  \bibinfo{author}{\bibfnamefont{J.}~\bibnamefont{Rührig}},
  \bibinfo{author}{\bibfnamefont{T.}~\bibnamefont{Pfau}}, \bibnamefont{and}
  \bibinfo{author}{\bibfnamefont{A.}~\bibnamefont{Griesmaier}},
  \bibinfo{journal}{Phys.~Rev.~A} \textbf{\bibinfo{volume}{89}},
  \bibinfo{pages}{043417} (\bibinfo{year}{2014}).

\bibitem[{\citenamefont{Ho and Zhou}(2009{\natexlab{a}})}]{ho_universal_2009}
\bibinfo{author}{\bibfnamefont{T.-L.} \bibnamefont{Ho}} \bibnamefont{and}
  \bibinfo{author}{\bibfnamefont{Q.}~\bibnamefont{Zhou}},
  \bibinfo{journal}{{arXiv}:0911.5506 [cond-mat]}
  (\bibinfo{year}{2009}{\natexlab{a}}), \bibinfo{note}{{arXiv}: 0911.5506}.

\bibitem[{\citenamefont{Ho and Zhou}(2009{\natexlab{b}})}]{ho_squeezing_2009}
\bibinfo{author}{\bibfnamefont{T.-L.} \bibnamefont{Ho}} \bibnamefont{and}
  \bibinfo{author}{\bibfnamefont{Q.}~\bibnamefont{Zhou}},
  \bibinfo{journal}{Proc.~Natl.~Acad.~Sci.} \textbf{\bibinfo{volume}{106}},
  \bibinfo{pages}{6916} (\bibinfo{year}{2009}{\natexlab{b}}).

\bibitem[{\citenamefont{Bernier et~al.}(2009)\citenamefont{Bernier, Kollath,
  Georges, De~Leo, Gerbier, Salomon, and Köhl}}]{bernier_cooling_2009}
\bibinfo{author}{\bibfnamefont{J.-S.} \bibnamefont{Bernier}},
  \bibinfo{author}{\bibfnamefont{C.}~\bibnamefont{Kollath}},
  \bibinfo{author}{\bibfnamefont{A.}~\bibnamefont{Georges}},
  \bibinfo{author}{\bibfnamefont{L.}~\bibnamefont{De~Leo}},
  \bibinfo{author}{\bibfnamefont{F.}~\bibnamefont{Gerbier}},
  \bibinfo{author}{\bibfnamefont{C.}~\bibnamefont{Salomon}}, \bibnamefont{and}
  \bibinfo{author}{\bibfnamefont{M.}~\bibnamefont{Köhl}},
  \bibinfo{journal}{Phys.~Rev.~A} \textbf{\bibinfo{volume}{79}},
  \bibinfo{pages}{061601} (\bibinfo{year}{2009}).

\bibitem[{\citenamefont{Heidrich-Meisner
  et~al.}(2009)\citenamefont{Heidrich-Meisner, Manmana, Rigol, Muramatsu,
  Feiguin, and Dagotto}}]{heidrich-meisner_quantum_2009}
\bibinfo{author}{\bibfnamefont{F.}~\bibnamefont{Heidrich-Meisner}},
  \bibinfo{author}{\bibfnamefont{S.~R.} \bibnamefont{Manmana}},
  \bibinfo{author}{\bibfnamefont{M.}~\bibnamefont{Rigol}},
  \bibinfo{author}{\bibfnamefont{A.}~\bibnamefont{Muramatsu}},
  \bibinfo{author}{\bibfnamefont{A.~E.} \bibnamefont{Feiguin}},
  \bibnamefont{and} \bibinfo{author}{\bibfnamefont{E.}~\bibnamefont{Dagotto}},
  \bibinfo{journal}{Phys.~Rev.~A} \textbf{\bibinfo{volume}{80}},
  \bibinfo{pages}{041603} (\bibinfo{year}{2009}).

\bibitem[{\citenamefont{Shuman et~al.}(2010)\citenamefont{Shuman, Barry, and
  DeMille}}]{shuman_laser_2010}
\bibinfo{author}{\bibfnamefont{E.~S.} \bibnamefont{Shuman}},
  \bibinfo{author}{\bibfnamefont{J.~F.} \bibnamefont{Barry}}, \bibnamefont{and}
  \bibinfo{author}{\bibfnamefont{D.}~\bibnamefont{DeMille}},
  \bibinfo{journal}{Nature Phys.} \textbf{\bibinfo{volume}{467}},
  \bibinfo{pages}{820} (\bibinfo{year}{2010}).

\bibitem[{\citenamefont{Zeppenfeld et~al.}(2012)\citenamefont{Zeppenfeld,
  Englert, Glöckner, Prehn, Mielenz, Sommer, van Buuren, Motsch, and
  Rempe}}]{zeppenfeld_sisyphus_2012}
\bibinfo{author}{\bibfnamefont{M.}~\bibnamefont{Zeppenfeld}},
  \bibinfo{author}{\bibfnamefont{B.~G.~U.} \bibnamefont{Englert}},
  \bibinfo{author}{\bibfnamefont{R.}~\bibnamefont{Glöckner}},
  \bibinfo{author}{\bibfnamefont{A.}~\bibnamefont{Prehn}},
  \bibinfo{author}{\bibfnamefont{M.}~\bibnamefont{Mielenz}},
  \bibinfo{author}{\bibfnamefont{C.}~\bibnamefont{Sommer}},
  \bibinfo{author}{\bibfnamefont{L.~D.} \bibnamefont{van Buuren}},
  \bibinfo{author}{\bibfnamefont{M.}~\bibnamefont{Motsch}}, \bibnamefont{and}
  \bibinfo{author}{\bibfnamefont{G.}~\bibnamefont{Rempe}},
  \bibinfo{journal}{Nature Phys.} \textbf{\bibinfo{volume}{491}},
  \bibinfo{pages}{570} (\bibinfo{year}{2012}).

\bibitem[{\citenamefont{Mathey et~al.}(2009)\citenamefont{Mathey, Tiesinga,
  Julienne, and Clark}}]{mathey}
\bibinfo{author}{\bibfnamefont{L.}~\bibnamefont{Mathey}},
  \bibinfo{author}{\bibfnamefont{E.}~\bibnamefont{Tiesinga}},
  \bibinfo{author}{\bibfnamefont{P.~S.} \bibnamefont{Julienne}},
  \bibnamefont{and} \bibinfo{author}{\bibfnamefont{C.~W.} \bibnamefont{Clark}},
  \bibinfo{journal}{Phys.~Rev.~A} \textbf{\bibinfo{volume}{80}},
  \bibinfo{pages}{030702} (\bibinfo{year}{2009}).

\bibitem[{\citenamefont{Tiesinga et~al.}(1993)\citenamefont{Tiesinga, Verhaar,
  and Stoof}}]{tiesinga_1993}
\bibinfo{author}{\bibfnamefont{E.}~\bibnamefont{Tiesinga}},
  \bibinfo{author}{\bibfnamefont{B.~J.} \bibnamefont{Verhaar}},
  \bibnamefont{and} \bibinfo{author}{\bibfnamefont{H.~T.~C.}
  \bibnamefont{Stoof}}, \bibinfo{journal}{Phys. Rev. A}
  \textbf{\bibinfo{volume}{47}}, \bibinfo{pages}{4114} (\bibinfo{year}{1993}).

\bibitem[{\citenamefont{Chin et~al.}(2010)\citenamefont{Chin, Grimm, Julienne,
  and Tiesinga}}]{feshbach_1}
\bibinfo{author}{\bibfnamefont{C.}~\bibnamefont{Chin}},
  \bibinfo{author}{\bibfnamefont{R.}~\bibnamefont{Grimm}},
  \bibinfo{author}{\bibfnamefont{P.}~\bibnamefont{Julienne}}, \bibnamefont{and}
  \bibinfo{author}{\bibfnamefont{E.}~\bibnamefont{Tiesinga}},
  \bibinfo{journal}{Rev. Mod. Phys.} \textbf{\bibinfo{volume}{82}},
  \bibinfo{pages}{1225} (\bibinfo{year}{2010}).

\bibitem[{\citenamefont{Gaebler et~al.}(2007)\citenamefont{Gaebler, Stewart,
  Bohn, and Jin}}]{gaebler_2007}
\bibinfo{author}{\bibfnamefont{J.~P.} \bibnamefont{Gaebler}},
  \bibinfo{author}{\bibfnamefont{J.~T.} \bibnamefont{Stewart}},
  \bibinfo{author}{\bibfnamefont{J.~L.} \bibnamefont{Bohn}}, \bibnamefont{and}
  \bibinfo{author}{\bibfnamefont{D.~S.} \bibnamefont{Jin}},
  \bibinfo{journal}{Phys. Rev. Lett.} \textbf{\bibinfo{volume}{98}},
  \bibinfo{pages}{200403} (\bibinfo{year}{2007}).

\bibitem[{\citenamefont{Kielhöfer}(2010)}]{kielhoefer}
\bibinfo{author}{\bibfnamefont{H.}~\bibnamefont{Kielhöfer}},
  \emph{\bibinfo{title}{{Variationsrechnung}}}
  (\bibinfo{publisher}{Vieweg+Teubner}, \bibinfo{year}{2010}),
  \bibinfo{edition}{1st} ed.

\end{thebibliography}

\end{document}